\documentclass[sigconf]{acmart}
\AtBeginDocument{%
  \providecommand\BibTeX{{%
    \normalfont B\kern-0.5em{\scshape i\kern-0.25em b}\kern-0.8em\TeX}}}

\copyrightyear{2024}
\acmYear{2024}
\setcopyright{rightsretained}
\acmConference[ICCAD '24]{IEEE/ACM International Conference on Computer-Aided Design}{October 27--31, 2024}{New York, NY, USA}
\acmBooktitle{IEEE/ACM International Conference on Computer-Aided Design (ICCAD '24), October 27--31, 2024, New York, NY, USA}
\acmDOI{10.1145/3676536.3676728}
\acmISBN{979-8-4007-1077-3/24/10}
\usepackage[inline]{enumitem}
\usepackage{xcolor}
\usepackage{pifont}
\usepackage{makecell}
\usepackage{threeparttable}
\usepackage{multirow}
\usepackage{mathtools}
\usepackage{subfig}
\usepackage[super]{nth}

\usepackage[  round-mode = places, round-precision = 2]{siunitx}
\definecolor{green}{RGB}{0,128,0}
\definecolor{red}{RGB}{255,0,0}
\newcommand{\greentick}{\textcolor{green}{\ding{51}}}
\newcommand{\redx}{\textcolor{red}{\ding{55}}}
\newcommand{\low}{\textcolor{red}{\textbf{low}}}
\newcommand{\moderate}{\textcolor{orange}{\textbf{moderate}}}
\newcommand{\high}{\textcolor{green}{\textbf{high}}}
\newcommand{\lowgreen}{\textcolor{green}{\textbf{low}}}
\newcommand{\highred}{\textcolor{red}{\textbf{high}}}

\newcommand{\repolink}{\url{https://github.com/ehw-fit/approximate-popcount}}

\begin{document}

\title[Evolutionary Approximation of Ternary Neurons for On-sensor Printed Neural Networks] {Evolutionary Approximation of Ternary Neurons \\for On-sensor Printed Neural Networks} 

\author{Vojtech Mrazek}
\affiliation{%
  \institution{Brno University of Technology}
  \city{Brno}
  \country{Czech Republic}}
\email{mrazek@fit.vutbr.cz}

\author{Argyris Kokkinis}
\affiliation{%
  \institution{Aristotle University of Thessaloniki}
  \city{Thessaloniki}
  \country{Greece}
}
\email{arkokkin@auth.gr}

\author{Panagiotis Papanikolaou}
\affiliation{%
  \institution{University of Michigan}
  \city{Ann Arbor}
  \country{USA}
}
\email{panagip@umich.edu }

\author{Zdenek Vasicek}
\affiliation{%
  \institution{Brno University of Technology}
  \city{Brno}
  \country{Czech Republic}
}
\email{vasicek@fit.vutbr.cz}

\author{Kostas Siozios}
\affiliation{%
  \institution{Aristotle University of Thessaloniki}
  \city{Thessaloniki}
  \country{Greece}
}
\email{ksiop@auth.gr}

\author{Georgios Tzimpragos}
\affiliation{%
  \institution{University of Michigan}
  \city{Ann Arbor}
  \country{USA}
}
\email{gtzimpra@umich.edu}

\author{Mehdi Tahoori}
\affiliation{%
  \institution{Karlsruhe Institute of Technology}
  \city{Karlsruhe}
  \country{Germany}
}
\email{mehdi.tahoori@kit.edu}

\author{Georgios Zervakis}
\affiliation{%
  \institution{University of Patras}
  \city{Patras}
  \country{Greece}
}
\email{zervakis@upatras.gr}

\renewcommand{\shortauthors}{V. Mrazek, A. Kokkinis, P. Papanikolaou, Z. Vasicek, K. Siozios, G. Tzimpragos, M. Tahoori, and G. Zervakis}

\begin{abstract}
Printed electronics offer ultra-low manufacturing costs and the potential for on-demand fabrication of flexible hardware.
However, significant intrinsic constraints stemming from their large feature sizes and low integration density pose design challenges that hinder their practicality.
In this work, we conduct a holistic exploration of printed neural network accelerators, starting from the analog-to-digital interface---a major area and power sink for sensor processing applications---and extending to networks of ternary neurons and their implementation.  
We propose bespoke ternary neural networks using approximate popcount and popcount-compare units, developed through a multi-phase evolutionary optimization approach and interfaced with sensors via customizable analog-to-binary converters.
Our evaluation results show that the presented designs outperform the state of the art, achieving at least  6$\times$ improvement in area and 19$\times$ in power. To our knowledge, they represent the first open-source digital printed neural network classifiers capable of operating with existing printed energy harvesters.
\end{abstract}

\begin{CCSXML}
<ccs2012>
   <concept>
       <concept_id>10010583.10010786</concept_id>
       <concept_desc>Hardware~Emerging technologies</concept_desc>
       <concept_significance>500</concept_significance>
       </concept>
   <concept>
       <concept_id>10010583.10010600.10010615</concept_id>
       <concept_desc>Hardware~Logic circuits</concept_desc>
       <concept_significance>500</concept_significance>
       </concept>
   <concept>
       <concept_id>10010147.10010257</concept_id>
       <concept_desc>Computing methodologies~Machine learning</concept_desc>
       <concept_significance>500</concept_significance>
       </concept>
 </ccs2012>
\end{CCSXML}
\ccsdesc[500]{Hardware~Emerging technologies}
\ccsdesc[500]{Hardware~Logic circuits}
\ccsdesc[500]{Computing methodologies~Machine learning}

\keywords{Approximate Computing, Electrolyte-gated FET, Printed Electronics, Low-Power Classifiers, Ternary Neural Networks}


\maketitle

\section{Introduction}\label{sec:intro}
Printed electronics hold significant promise for various applications, including smart packaging~\cite{smartpackaging2022}, in-situ monitoring~\cite{monitoring:EPTC:2017}, forensics~\cite{salivary:Talanta:2020}, and accessible healthcare products and wearables~\cite{bodytemperature:sna:2020,pressuresensor:research:2022,wearable:adma:2022,Wearable:acssensors:2019,healthcare:Nanoscale:2024}.
A common denominator among these applications is the need for devices that are stretchable, porous, flexible, and conformal, all while adhering to strict area and power budgets.
Another common feature is that processing typically involves classification, with recent research efforts directed towards the realization of printed bespoke machine learning (PBML) accelerators~\cite{Mubarik:MICRO:2020:printedml, Iordanou:Nature:2024}.

The term bespoke denotes fully customized circuit implementations (e.g., hardwired model parameters~\cite{Mubarik:MICRO:2020:printedml,Ozer:2019:Bespoke,Ozer:Nature:2020}) as well as approximations~\cite{Armeniakos:TCAD2023:cross,Armeniakos:TC2023:codesign,Afentaki:ICCAD2023:axmac,Afentaki:DATE2024:gatrain,Balaskas:ISQED2022:axDT, Kokkinis:DATE2023} per ML model and dataset.
While this assumption may seem weak in other contexts%(perhaps with the exception of FPGA-based solutions)
, it is well-suited to printed electronics, where devices are considered disposable~\cite{disposable:JSNB:2023} due to their low fabrication cost and short turnaround times~\cite{cui2016printed, chang2017circuits, Mubarik:MICRO:2020:printedml}.
This specialization allows existing PBML designs, at least in theory, to operate with the power provided by available printed batteries~\cite{PrintedBatteries2018}.
However, the practicality of these solutions is often compromised by the overlooked cost of sensor-processor interfacing, a crucial aspect of the overall design since processing occurs directly on sensor data. 

In this work, we address this problem holistically, starting our optimizations from the sensor boundary. Firstly, we replace costly analog-to-digital converters (ADCs) with efficient customized analog-to-binary converters (ABCs).
Secondly, we explore ternary neural networks (TNNs)~\cite{TNN}, where weights take only three values: $\{-1, 0, 1\}$, thus eliminating the need for multipliers.
Thirdly, to further enhance hardware efficiency, we introduce  approximate popcount and fused popcount-and-compare units.
To facilitate their automated utilization, we implement a set of evolutionary multi-objective optimization techniques.

The result is approximate bespoke TNN designs that, on average, use $41$\% fewer resources and consume $42$\% less power than their exact counterparts, with no accuracy loss on common printed ML datasets~\cite{Mubarik:MICRO:2020:printedml, Weller:2021:printed_stoch}.
Compared to state-of-the-art digital approximate printed classifiers, our evaluation indicates an average reduction of 32$\times$ in area and 34$\times$ in power consumption, when factoring in the gains from replacing ADCs with the proposed ABC design.

In summary, the main \textbf{contributions} of this research are:
\begin{itemize}[leftmargin=12pt]
    \item A custom analog-to-binary converter design that uses only two resistors and an analog comparator.
    \item Approximate designs for popcount and fused popcount-and-comparison units, accompanied by a detailed Pareto analysis.
    \item The development of an multi-phase evolutionary optimization framework for the automated deployment of our approximation techniques and the search of the formed design space.
    \item The first, to the best of our knowledge, open-source complete digital printed classifier solution\footnote{Available at: \repolink} that can be fully powered by either available printed batteries~\cite{PrintedBatteries2018} or energy harvesters~\cite{printedharvester}.
\end{itemize}

\section{Related Work}
Printed electronics frequently targets applications centered on classification tasks.
This focus, combined with the expanding fast-moving consumer goods market, has spurred research into printed machine learning accelerators.
Various efforts have explored neural network models in this context, as summarized in Table~\ref{tab:motsoa}.

The design choices behind the proposed solutions are shaped by the technology's unique attributes and the requirements imposed by its application domains.
Printed electronics employ additive and mask-less manufacturing techniques like jet, screen, or gravure printing~\cite{cui2016printed}.
These methods, coupled with low capital equipment costs, enable rapid and inexpensive circuit production---sometimes for less than a cent per unit. 
However, these same characteristics lead to large feature sizes, low integration density (far below silicon VLSI), and increased device latencies~\cite{lei2019low, cadilha2017digital}.
The latter is often not a concern due to the low sampling rates and relaxed performance requirements in typical applications~\cite{Henkel:ICCAD2022:expedition}.
As for the rest, to maximize functionality with minimum resources, bespoke designs are favored, especially since circuit programmability and reusability are less critical in this domain.

Bespoke designs are customized for specific classifier models trained on particular datasets, eliminating generality-related overheads~\cite{Kumar2017Bespoke}.
This specialization extends to developing and implementing tailored approximation techniques~\cite{Han:ETS2013}.
For instance, Armeniakos et al.~\cite{Armeniakos:DATE2022:axml} proposed post-training weight replacement with more hardware-friendly approximates to reduce the cost of the underlying multipliers.
This effort was further enhanced by introducing voltage over-scaling techniques where necessary~\cite{Armeniakos:TCAD2023:cross}.
In subsequent works, Armeniakos et al.~\cite{Armeniakos:TC2023:codesign} incorporated hardware-friendly weight approximation into the training process and implemented post-training addition approximation using simple truncation.
Afentaki et al.~\cite{Afentaki:ICCAD2023:axmac, Afentaki:DATE2024:gatrain} constrained weights to powers~of~2---leveraging that, in bespoke circuits, a multiplication by a power~of~2 is implemented simply by rewiring---approximated accumulations by pruning the adder trees, and employed bounded low-precision ReLu activation.
Lastly, an alternative approximation method, stochastic computing~\cite{Liu:TNNLS2021} neural networks, was explored by Weller et al.~\cite{Weller:2021:printed_stoch}.

The work presented here distinguishes itself from existing literature in several ways: it avoids the potential for significant accuracy losses associated with stochastic computing approaches; it employs a multiplier-free design and approximate adders for maximum efficiency; it is purely digital, avoiding the noise and variability issues that can affect low-resolution printed analog designs~\cite{Zhao:DATE2023,Zhao:ICCAD2022,Zhao:ICCAD2023,Pal:DATE204}; and it begins optimization at the sensor boundary, rather than post-ADC.
To our knowledge, the only other printed classifier that considers sensor-processor interface optimizations is a unary decision tree accelerator~\cite{ Armeniakos:DATE2024:adcdt}.
However, this design relies on multiple analog comparators per analog-to-unary converter, indicating room for further improvement.

\begin{table}[t]
  \centering
  \small
  \setlength\tabcolsep{3pt}
  \renewcommand{\arraystretch}{1.1}
  \caption{State-of-the-art printed neural networks.}\vspace{-2ex}
 \label{tab:motsoa}

      \begin{tabular}{|c|c|c|c|c|c|}
      \hline
      \textbf{Ref.} & \textbf{Digital} & \makecell{\textbf{Ax.} \\ \textbf{Mult.}} & \makecell{\textbf{Ax.} \\ \textbf{Add.}} & \makecell{\textbf{Low Prec.} \\ \textbf{Act. Func.}} & \makecell{\textbf{Inp. Size/} \\ \textbf{ADC Cost}} 
  \\ \hline    \hline

\cite{Mubarik:MICRO:2020:printedml} & \greentick & \redx  & \redx  & \redx & \highred \\ \hline 
\cite{Armeniakos:DATE2022:axml,Armeniakos:TCAD2023:cross} & \greentick & \low & \redx & \redx & \highred\\ \hline 
\cite{Armeniakos:TC2023:codesign} & \greentick & \moderate & \low &  \redx  &  \highred    \\ \hline 
\cite{Afentaki:ICCAD2023:axmac,Afentaki:DATE2024:gatrain}  & \greentick & \high & \moderate &  \greentick &  \highred \\ \hline 
\cite{Weller:2021:printed_stoch} & \redx & \redx  & \redx  & \greentick & n/a \\ \hline
\hline
     \textbf{Proposed} & \greentick & \high & \high & \greentick  & \lowgreen \\ \hline
      \end{tabular}
      \vspace{-2ex}
\end{table}

\section{Proposed End-to-End Design}\label{sec:tnn}

\subsection{Sensor-Processor Interface}
\begin{figure}[t]
    \centering
    \includegraphics[scale=0.35]{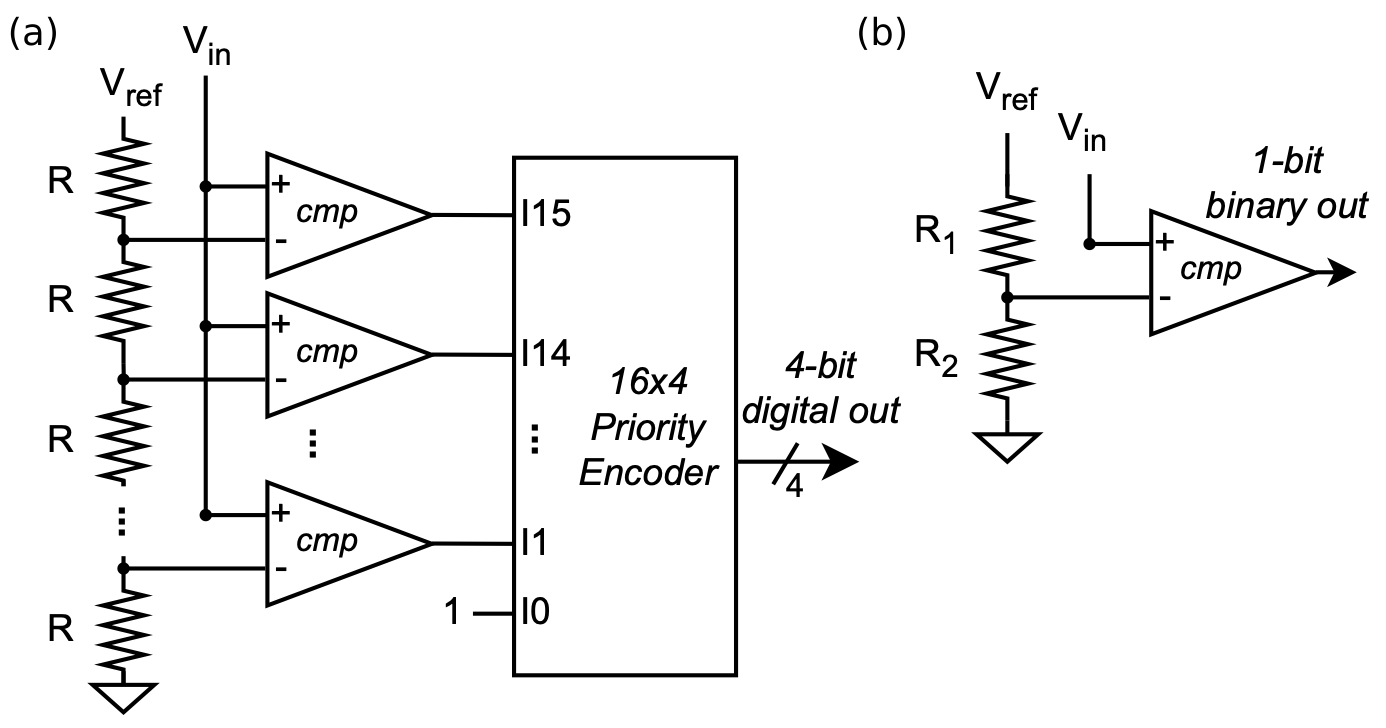}
    %\vspace{-2mm}
    \caption{Panel a: A $4$-bit flash ADC. Panel b: Proposed analog-to-binary converter. The ratio R$_1$/R$_2$ is used to regulate the voltage threshold at which an input feature becomes 1. $V_{in}$ denotes the voltage output of the sensor, which feeds the two converters (a) or (b).}
    \label{fig:adc}
    \vspace{-2ex}
\end{figure}

The sensor-processor interface typically consists of an array of ADCs.
ADCs are often overlooked in existing optimization efforts despite their significant hardware costs.
For example, a 4-bit printed flash ADC, as the one shown in \autoref{fig:adc}a, occupies $12$~mm$^2$ and consumes $1$~mW of power~\cite{Armeniakos:DATE2024:adcdt}.
If incorporated into a recent printed multilayer perceptron design with 4-bit inputs~\cite{Afentaki:DATE2024:gatrain}, it increases its area by $1.11\times$ to $44\times$ and power consumption by $1.97\times$ to $370\times$.  

We propose to replace such ADCs with a custom analog-to-binary converter (ABC), presented in \autoref{fig:adc}b.
Compared to the reference $4$-bit ADC, our ABC design requires 8$\times$ fewer resistors, 15$\times$ less comparators, and does not require a priority encoder.
This translates to an area of only 0.07~mm$^2$ and a power consumption of 0.03~mW---167$\times$ smaller and 34$\times$ more power efficient than the original ADC design.
Additionally, we allow the use of resistors, with potentially different values $R_1$ and $R_2$.
Having more than one voltage rails is inefficient in printed electronics, so we assume a common reference voltage $V_{ref}$ for all ABCs. 
Adjusting the $R_1/R_2$ ratio enables to define the voltage threshold at which a specific input feature transitions between $1$ and $0$.

\subsection{Bespoke Ternary Neural Networks}
This works focuses on ternary neural networks.
TNNs provide accuracy levels that fall between binary and standard neural networks, are easy to implement, and eliminate the need for hardware multipliers.
Previous printed multilayer perceptron designs also avoided multipliers by constraining synaptic weight values to powers~of~2~\cite{Afentaki:ICCAD2023:axmac,Afentaki:DATE2024:gatrain, Kokkinis:DATE2023}.
However, this wider range of weight values can result in larger accumulation circuits compared to TNNs, where weights are restricted to only $-1$, $0$, or $1$.
For example, assuming $4$-bit inputs $I_j$, the sum $8\times I_1 - 2\times I_2 - 4\times I_3 - 2\times I_4$ requires 0.67~cm$^2$, while with ternary weights, the area of the respective sum is at worst $0.38$~cm$^2$, i.e., $43$\% smaller.
These improvements are enhanced under the assumption of binary inputs ($I_j \in \{0,1\}$), such as those provided by the proposed ABCs.
In this case, the corresponding area reduces to $0.2$~cm$^2$, i.e., $3.4$$\times$ smaller than the initial design with powers~of~2 weights.

\subsubsection{Inputs and Weights Quantization}
We employ Qkeras~\cite{Coelho:Qkeras} for quantization-aware training (QAT) of the TNN, allowing customization of quantizers for synaptic weights and inputs.
For weights, we use the \texttt{ternary} quantizer with \texttt{alpha} set to $1$, yielding coefficients in $\{-1, 0, 1\}$.
Hidden-layer neuron inputs are quantized to $\{0, 1\}$ using the \texttt{quantized\_bits} activation function.
For first-layer neuron inputs, we extend the \texttt{quantized\_bits} function to set specific quantization thresholds for each input feature.
We normalize training set inputs to $[0, 1]$ and analyze their distribution to adjust the quantization threshold $V_q$ based on whether the distribution is left-skewed, nearly normal, or right-skewed.
This threshold determines whether an input is considered $1$ or $0$ in the binary representation.

In this work, we set \(V_q\) to the median of the normalized input distribution for each input, rather than treating it as a learnable parameter during training. However, process variations can affect the actual values of R$_1$ and R$_2$, potentially altering the R$_1$/R$_2$ ratio in the ABC design. This makes \(V_q\) susceptible to process variations, which could negatively impact classification accuracy. Optimizing for this is beyond the scope of this paper, but we propose incorporating variation-aware training techniques into QAT~\cite{Zhao:DATE2023,Zhao:ICCAD2022} to enhance the robustness of our TNNs.

%While \(V_q\) could be incorporated as a learnable parameter during training, in this work, we set \(V_q\) based on each input's statistics, specifically by setting it equal to the median of the corresponding normalized input distribution. 

%Printed electronics are subject to considerable process variations, leading to significant fluctuations in the sizes of R$_1$ and R$_2$.
%These latter may alter the R$_1$/R$_2$ ratio in the ABC design, making \( V_q \) sensitive to process variations and negatively impacting the final classification accuracy.
%However, incorporating variation-aware training techniques, such as ~\cite{Zhao:DATE2023,Zhao:ICCAD2022}, into QAT can enhance the robustness of our TNNs against these variations.

\subsubsection{Circuit Implementation}
The trained TNN model is automatically translated into hardware by combining Icarus Verilog (iverilog) with a set of custom neuron templates designed for each layer.
Adhering to bespoke design principles, the synaptic weights are hardcoded into the circuit description.
For synaptic weights with a value of $0$ between the input and hidden layer, the corresponding connection is removed from the hidden-layer neuron, effectively eliminating an addend from its accumulator.
For non-zero weights, the sign determines the hardware operation---either addition or subtraction---over the respective input. 
Consequently, each hidden-layer neuron performs a multi-operand summation, with its output being the sign of the respective sum: 1 for a positive sum and 0 for a negative sum.

The output layer implementation is customized to accommodate the different encoding of hidden-layer neuron outputs ($\{-1, 1\}$) compared to ABC outputs ($\{0, 1\}$).
We employ XNOR-based neurons~\cite{Rastegari:ECCV2016:xnornet}, which use XNOR gates instead of multiplications and a popcount operation to determine the number of high inputs.

Given the fixed, known weights, XNOR gates are simplified: becoming NOT gates for weights of $-1$ and simple wires for weights of $1$. 
The popcount operation is performed on $\{0, 1\}$ values, while our encoding assumes $\{-1, 1\}$. Thus, we apply a linear mapping that translates $0$ to $\frac{1}{2}$. So, unlike the hidden layer connections, zero-valued weights are retained here as a $+0.5$ constant correction term.
By ensuring, in training, the output neurons have the same number of zero-valued connections $N$, the correction term is consistent across all neurons and can be omitted without affecting the result of the classification, as $\mathrm{argmax}(y+\frac{N}{2},q+\frac{N}{2}) = \mathrm{argmax}(y,q)$.

% Given the fixed, known weights, XNOR gates are simplified: becoming NOT gates for weights of $-1$ and simple wires for weights of $1$. 
% The popcount operation is performed on $\{0, 1\}$ values, while our encoding assumes $\{-1, 1\}$; thus, we apply a linear mapping $\frac{y+1}{2}$ from $\{-1, 1\}$ to $\{0, 1\}$.
% This transformation maps $0$ to $\frac{1}{2}$, so unlike the hidden layer connections, zero-valued weights are retained here as a $+0.5$ constant correction term.
% By ensuring, in training, the output neurons have the same number of zero-valued connections $N$ (i.e., same sparsity), the correction term is consistent across all neurons and can be omitted without affecting the result of the classification, as $\mathrm{argmax}(y+\frac{N}{2},q+\frac{N}{2}) = \mathrm{argmax}(y,q)$.

\begin{figure}[t]
    \centering
    \includegraphics[scale=0.99]{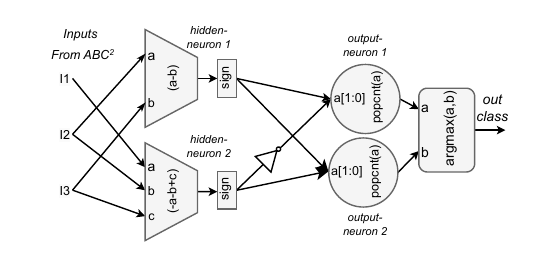}%\vspace{-1ex}
    \caption{Bespoke $(3,2,2)$ TNN circuit with $[[[0,1,-1],[-1,-1,1]], [[1,-1],[1,1]] ]$ weights.}
    \label{fig:TNN}
    \vspace{-2ex}
\end{figure}

Lastly, to implement the argmax function, which identifies the output neuron with the highest score (i.e., highest popcount output), a set of comparators is employed. \autoref{fig:TNN} depicts a bespoke (3,2,2) TNN design. In this illustration, the hidden-layer neurons have weights $[0,1,-1]$ and $[-1,-1,1]$, while the output neurons have weights $[1,-1]$ and $[1,1]$.

\section{Proposed Approximation}
\label{sec:ppc}

\begin{figure}[t]
    \centering
    \includegraphics[width=\columnwidth]{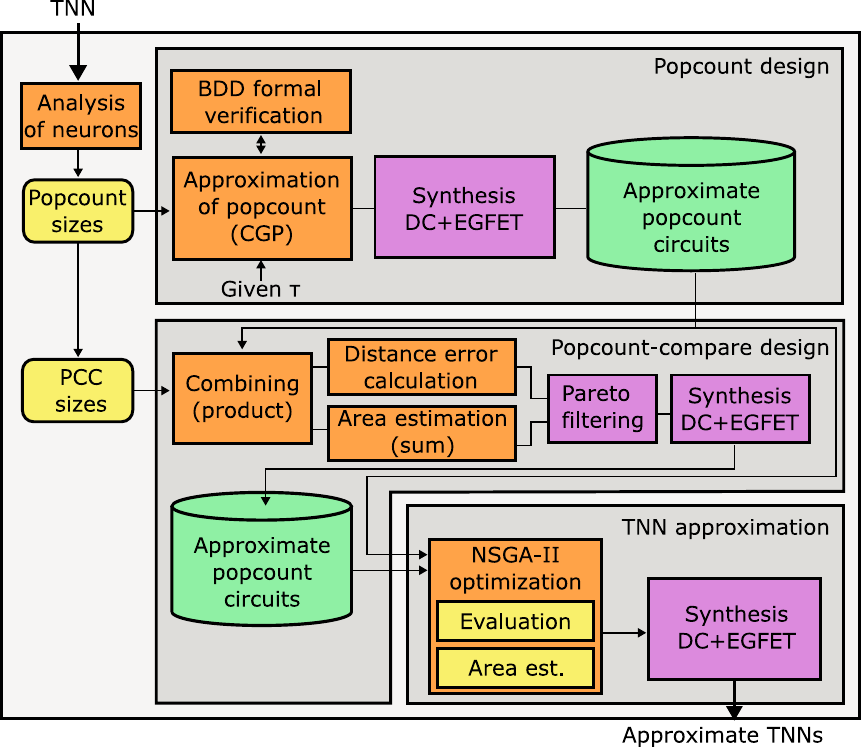}%\vspace{-1ex}
    \caption{Overview of the proposed three-phase TNN approximation framework, including Cartesian genetic programming for popcount circuit approximations in Phase 1, a Pareto analysis for identifying optimal combinations of popcount circuits for the construction of approximate popcount-compare circuits in Phase 2, and their integration into bespoke TNN circuits using the NSGA-II evolutionary algorithm in Phase 3. The underlying technology assumed is EGFET, operating at voltages as low as 0.6-1~V.}
    \label{fig:overall}\vspace{-2ex}
\end{figure}

The proposed approximation methodology, illustrated in \autoref{fig:overall}, consists of three main phases.
In Phase 1, Cartesian genetic programming (CGP)~\cite{miller:cgp} is used to evolve circuits that approximate the popcount function.
In Phase 2, we identify Pareto-optimal combinations of these circuits to deliver the necessary approximate popcount-compare functionality for the hidden-layer neurons.
Finally, in Phase 3, the NSGA-II evolutionary algorithm~\cite{deb:nsga} is applied to integrate these approximate components into a bespoke TNN circuit, optimizing for maximum resource efficiency with minimal impact on accuracy. 

%\subsection{Phases 1 \& 2 : Popcount (-compare) Circuit Approximations and Pareto Analysis}
% \subsection{Phases 1 \& 2 : Approximate Popcount \& Popcount-compare Circuits}
\subsection{Approximate Popcount \& Popcount-compare Circuits}
For hidden-layer neurons, a binary step activation function is used. The computation per neuron is:
\begin{equation}
\sum_{\forall i} w_i I_i \geq 0,
\end{equation}
where $w_i$ are synaptic weights and $I_i$ are inputs. Given $w_i \in \{-1, 0, 1\}$, this transforms to:
\begin{equation}
\sum_{\mathclap{\forall i: w_i = 1}}{I_i} \;-\; \sum_{\mathclap{\forall i: w_i = -1}}{I_i} \geq 0 
\implies
\sum_{\mathclap{\forall i: w_i = 1}}{I_i} \;\geq\; \sum_{\mathclap{\forall i: w_i = -1}}{I_i}.
\label{eq:neurontransformed}
\end{equation}

With $I_i \in \{0,1\}$, each sum can be calculated by a popcount operation. Thus, each hidden-layer neuron requires two popcount circuits and a comparator. The comparator's complexity is minimal compared to this of the popcount operations. For example, with 8 inputs, the comparator uses only 16\% of the hidden-layer neuron's total area; for 64 inputs, it's less than 3\%. In the rest of this Section, we introduce approximate implementations of popcount-compare (for hidden-layer neurons) and popcount (for output-layer neurons) circuits to enhance area-efficiency.

\subsubsection{Approximate Popcount Circuits}
An $n$-bit popcount (PC) operation sums $n$ binary inputs, typically implemented using a tree structure of adders.
To approximate this exact circuit, we employ the evolutionary CGP algorithm~\cite{miller:cgp}.
In CGP, a simple search method is deployed over an integer address-based genetic representation of the circuit to create random mutations of an initial population $P$ that contains the accurate PC circuit. To evaluate the candidate circuits, we use a fitness function.
Each member $c$ in population $P$ receives a fitness score $F(c)$, with the highest-scoring individual becoming the parent of the next population. This parent then generates $\lambda$ new candidate solutions through mutation.
The process terminates when either the number of iterations exceeds a set maximum, or a predefined time limit is reached.

The searching algorithm's optimization criterion is the estimated area for the target technology (e.g., EGFET PDK~\cite{Bleier:ISCA:2020:printedmicro}), while maintaining the error $\varepsilon$ below a specific threshold, consistent with previous error-oriented approximation research~\cite{mrazek:iccad:2016}.
This error constraint $\tau$ is incorporated into the fitness function as follows:
\begin{equation}   
    F(c) = \begin{cases}
        area(c), & \text{if} \varepsilon(c) \leq \tau \\
        \infty, & otherwise.
    \end{cases}
\end{equation}

PC circuits produce standard arithmetic outputs, allowing the use of conventional error metrics in the approximation process: mean arithmetic error $\varepsilon_{mae}$ and worst-case arithmetic error $\varepsilon_{wcae}$. These metrics calculate the average or maximum error across all input combinations. However, for a large number of inputs $n$ in PC circuits, evaluating all $2^n$ input vectors becomes impractical. To address this, a formal verification approach using binary decision diagrams (BDD)~\cite{soeken:bdd} is used.

\subsubsection{Approximate Popcount-compare Circuits}
Hidden-layer neurons are implemented using popcount-compare (PCC) circuits to realize Equation~\eqref{eq:neurontransformed}.
Each PCC consists of two PC circuits with $n_{pos}$ and $n_{neg}$ input bits, respectively, and a $j$-bit comparator, where $j$ is equal to $\lceil\log_2\max{(n_{pos}, n_{neg})}\rceil$. To ensure a comprehensive analysis and evaluation, PCCs are created for all $n_{pos}$ and $n_{neg}$ configurations found in the target TNNs.\smallskip

\noindent\textbf{Distance metric}: 
In approximating non-arithmetic circuits, Hamming distance is often considered the sole suitable error metric.
However, this approach falls short in the context of approximate computing, as it may not accurately represent the error levels~\cite{approxerrors}.
For example, in $4$-bit output arithmetic circuits, an error in one bit position ($0 \rightarrow 8$) can be more severe than errors in all four bits ($7 \rightarrow 8$).
This limitation extends to relational operators as well.
Consider two approximate $\geq_a$ and $\geq_b$ circuits:  if $\geq_a$ erroneously determines that $0 \geq_a 1 = \text{TRUE}$ (a $1$-position deviation) and $\geq_b$ incorrectly asserts that $0 \geq_b 4 = \text{TRUE}$ (a $4$-position deviation), both errors result in the same Hamming distance of 1 (FALSE $\rightarrow$ TRUE), despite their clearly different magnitude of errors. 

The Hamming distance's inability to capture error severity in the context of approximate computing is of particular importance to PCC approximation, which produces a one-bit output.
To address this, we propose a new \textit{distance metric} $D$, inspired by those used in approximate median filters and sorting networks~\cite{mrazek:patmos}.
This metric is defined as:

\begin{equation}
    \label{eq:distance}
D(x,z) =\begin{cases}
			0, & \text{if $ rel(x,z) = rel'(x,z)$}\\
            x-z, & \text{otherwise}
		 \end{cases}
\end{equation}

The distance $D$ is a function of two inputs $x$ and $z$, an accurate relational function $rel$ with a Boolean output, and its approximate variant $rel'$.
This provides a more nuanced measure of error magnitude, addressing the limitations of Hamming distance in this context.
Using $D$, we define the mean distance error ($\varepsilon_{mde}$) and worst-case distance error ($\varepsilon_{wcde})$ as follows:
\begin{equation}
\varepsilon_{mde} = \frac{1}{|G|} \sum_{\forall (x,z) \in G}{|D(x, z)|}; \;
\varepsilon_{wcde} = \max_{\forall (x,z) \in G}{|D(x, z)|}
\label{eq:pccerror}
\end{equation}
Here, $G$ represents the input domain.\\\vspace{-2mm}

\noindent\textbf{Pareto analysis}: We create a library of approximate PCC circuits using the PC circuit approximation methodology described above. For each exact PCC circuit, we first generate all possible approximate PCC circuits by testing all combinations of $n_{pos}$ and $n_{neg}$ PC circuits from our approximate PC library, including exact PC circuits as zero-error designs. Subsequently, we identify Pareto-optimal PCC circuits using two criteria: (i)~accuracy, evaluated through $\varepsilon_{mde}$ for $10^6$ random ($x,z$) pairs in Equations \eqref{eq:distance} and \eqref{eq:pccerror}, and (ii) hardware cost, estimated by the combined area of the $n_{pos}$ and $n_{neg}$ PC circuits. 

This search process is fully parallelizable and can be conducted at a high level (i.e., in Python) without the need for hardware evaluation. Consequently, the execution time remains well constrained, ensuring the scalability of our approximate PCC circuit search.
Once the Pareto analysis is completed, the selected PCC circuits, including both comparators and approximate PC circuits, are synthesized to asses their actual hardware cost. 

%\subsection{Phase 3: Approximate Circuit Integration}
\subsection{Approximate Circuit Integration}
The last step is the integration of PC and PCC circuits from the developed libraries into a complete TNN circuit. 
The objective is to achieve the best area-accuracy trade-off.
To this end, we deploy a multi-objective optimization based on the evolutionary NSGA-II algorithm~\cite{deb:nsga}.
As in prior optimizations, the two criteria are: accuracy and area.
We encode the problem as an integer list, where each integer corresponds to one circuit from the PCC and PC library for the respective neuron.
The sum of the areas of the selected PCC and PC circuits is used as a proxy for the overall area of the TNN.
To evaluate the actual area and power results of the generated TNNs, we go again through a round of synthesis.
The underlying technology assumed is EGFET, which allows operation at voltages as low as 0.6-1~V~\cite{Marques:Materials:2019} and aligns well with low-power requirements of battery/self-powered IoT applications.

\section{Evaluation}\label{sec:eval}
The goal of the presented research is to develop bespoke TNN circuit designs that optimize hardware efficiency while minimizing any decrease in accuracy.
Initially, we evaluate the proposed hardware implementations and the efficacy of our approximation methods. Subsequently, we conduct a comparative analysis against the current state-of-the-art printed neural networks.\smallskip

%\noindent\textbf{Evaluation methodology \& assumptions:}
\noindent\textbf{Evaluation setup:}
For our evaluation, we utilize the five datasets from the UCI ML repository~\cite{uci}. These datasets are chosen for two primary reasons: they enable direct comparisons with state-of-the-art designs, and they consist of sensor data suitable for printed electronics' applications, as discussed in studies~\cite{Mubarik:MICRO:2020:printedml,Weller:2021:printed_stoch}.
To perform digital circuit synthesis and analysis, we rely on Synopsys Design Compiler and PrimeTime tools, used in conjunction with the EGFET standard cell library~\cite{Bleier:ISCA:2020:printedmicro}.
For the ABC design, we utilize Cadence Virtuoso with the EGFET PDK, running SPICE simulations to assess its area and power consumption.
The voltage supply for all circuits is set to 0.6~V, in line with EGFET capabilities and consistent with prior work in the field~\cite{Afentaki:ICCAD2023:axmac,Afentaki:DATE2024:gatrain}. In terms of operating speeds, all circuits are evaluated at a frequency of 5~Hz, although our designs can achieve frequencies up to 20~Hz. This decision is made to maintain consistency with existing solutions in the literature~\cite{Bleier:ISCA:2020:printedmicro,Armeniakos:TC2023:codesign} and facilitate a fair comparison.\smallskip

\noindent\textbf{TNN baseline:} Exact TNN models are used as the baseline for accuracy comparisons.
The selected datasets are split into $70$\% for training and $30$\% for testing.
For training, the Adam optimizer is used with the number of epochs ranging from $10$ to $20$.
The learning rate for each network is obtained through Bayesian optimization, aiming to maximize the model's inference accuracy with a maximum of $100$ attempts.
Learning rate parameters are selected from the range of $0.001$ to $0.01$.
The final model is that with the highest accuracy among a set of $100$ TNNs with various hyperparameters.

Regarding TNN hyperparameters, all models assume a single hidden layer.
To determine the number of hidden-layer neurons, we perform an exhaustive search within the range of $1$ to $40$ neurons.
For each iteration, we repeat the procedure described previously.
Among the TNNs yielding the highest accuracy, we select the one with the fewest neurons.
This exhaustive search takes less than one hour on an AMD EPYC 7552 server with 256~GB RAM.

Table~\ref{tab:table_tnnacc} summarizes our findings across the examined datasets.
For reference, we also include the accuracy numbers achieved by a baseline MLP network~\cite{Mubarik:MICRO:2020:printedml} utilizing $4$-bit inputs and $8$-bit weights.
The comparison indicates $0$-$4$\% accuracy drop for our baseline TNNs.
However, as will discussed in the rest of this section, this compromise enables a significant boost in hardware efficiency, ensuring that the operating requirements of our printed TNNs fall within the integration and power budget of the target technology. 

\begin{table}[t]
  \centering
  %\footnotesize
  \small 
  \setlength\tabcolsep{4pt}
  \caption{TNN accuracy results.}\vspace{-2ex}
  \label{tab:table_tnnacc}
  \begin{threeparttable}
    \begin{tabular}{|l|ccc|ccc|}
      \hline
     \multirow{2}{*}{\textbf{Dataset}}  & \multicolumn{3}{c|}{\textbf{Exact MLP~\cite{Mubarik:MICRO:2020:printedml}}} & \multicolumn{3}{c|}{\textbf{Our Exact TNN}} \\ \cline{2-7}
      & \textbf{A$^\star$} & \textbf{T$^\dagger$} & \textbf{I/W$^*$} 
      & \textbf{A$^\star$} & \textbf{T$^\dagger$} & \textbf{I/W$^*$} \\
      \hline     
      Arrhythmia & 0.62 & (274,5,16) & 4/8 & 0.60 & (274,3,16) & 1/2  \\ 
      \hline
       Breast Cancer & 0.98 & (10,3,2) & 4/8 & 0.98 & (10,10,2) & 1/2  \\ 
      \hline
      Cardio & 0.88 & (21,3,3) & 4/8 & 0.85 & (21,3,3) & 1/2  \\ 
      \hline 
       Redwine & 0.56 & (11,2,6) & 4/8 & 0.56 & (11,3,6) & 1/2 \\ 
      \hline
      Whitewine & 0.54 & (11,4,7) & 4/8 & 0.50 & (11,11,7) & 1/2 \\ 
      \hline
    \end{tabular}
    \begin{tablenotes}\footnotesize
      \item[] $^\star$Inference accuracy. $^\dagger$Network Topology. $^*$Bit-width Inputs/Weights.
    \end{tablenotes}
  \end{threeparttable}%\vspace{-4ex}
\end{table}

\subsection{Approximate Circuits Evaluation}
\subsubsection{Popcount Circuits} We generate approximate PC circuits with the following constraints and configurations.
The error limits $\tau_{mae}$ and $\tau_{wcae}$ are logarithmically distributed from $0.1$ to $0.5\cdot2^m$ and from $1$ to $0.5\cdot2^m$, with $m = \lceil \log_2 n \rceil$, resulting in $2,\!090$ approximate circuits.
We also set CGP search termination criteria to $30$, $60$, and $300$ minutes for PC sizes $n<16$, $n<32$, and $n<60$.
Using BDD evaluation, we achieve average speeds of $7,\!759$ and $1,\!362$ evaluations per second for $16$-bit and $32$-bit PCs.
For $60$-bit PC circuits, this number falls to $25$ evaluations per second.
To account for this, we extend the time limit, thereby increasing the chances of identifying solutions with better area-accuracy trade-offs.

\begin{figure}[t]
    \centering
    \includegraphics[width=\columnwidth]{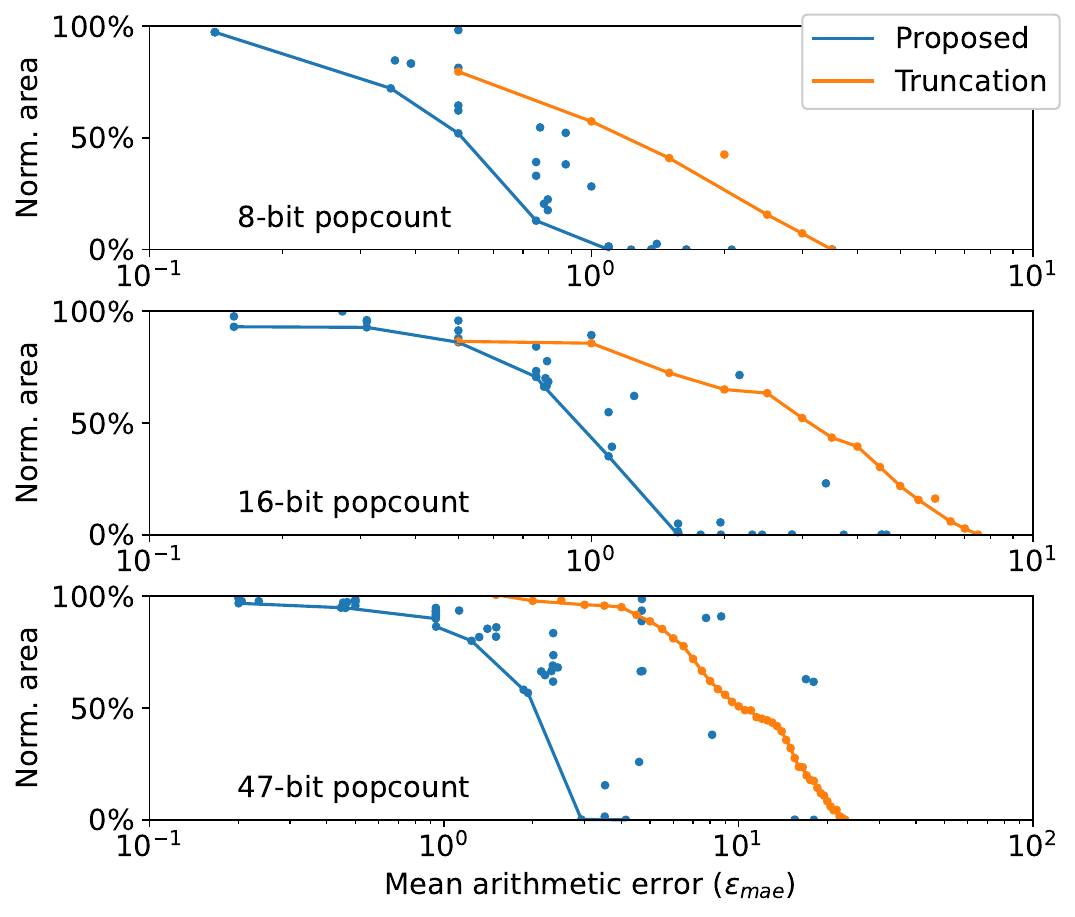}\vspace{-1em}
    \caption{Comparison between the proposed approximation approach and the previously used truncation technique for PCC circuits of various sizes. The results displayed are based on post-synthesis area measurements.}%\vspace{-4ex}
    \label{fig:pc_selected}
\end{figure}

A visualization of the obtained results for three PC circuit sizes is provided in \autoref{fig:pc_selected}.
The featured circuits are synthesized and evaluated based the error metrics and methodology described above.
For easier comparison, area results are normalized relative to exact PC circuit of the same size. Our results indicate that as the error limit increases, CGP optimization identifies increasingly area-efficient solutions.
Additionally, we assess the efficacy of our approximation approach through a comparison of our PC circuits with variants derived using the truncation approximation approach, which has previously been applied in a wide range of approximate circuits, including multilayer perceptrons~\cite{venkataramani:axnn}.
In this case, we achieve about $2\times$ area reduction with $\varepsilon_{mae}$ of only $0.5$ for an $8$-bit PC circuit, $1.1$ for a $16$-bit PC circuit, and $1.9$ for a $47$-bit PC circuit.
Similar results to those shown in \autoref{fig:pc_selected} are observed for all PC circuit sizes.

\subsubsection{Popcount-compare Circuits}

\begin{figure}[h]
    \subfloat{\label{fig:pcc_estimated}}
    \subfloat{\label{fig:pcc_dist}}
    \centering
    \includegraphics[width=\columnwidth]{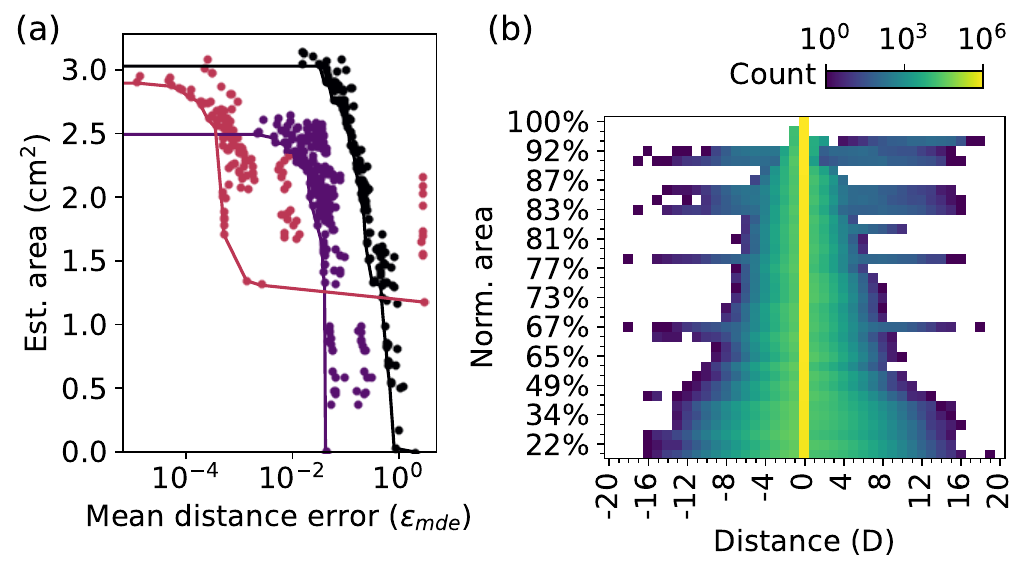}\vspace{-1.3em}
    \caption{Panel a: Trade-off illustration for three PCC circuits applied to the Arrhythmia dataset. The circuits are characterized by their sizes $\boldsymbol{(n_{pos}, n_{neg})}$: black $\boldsymbol{(45, 39)}$, purple $\boldsymbol{(47, 30)}$, and pink $\boldsymbol{(60, 29)}$. Panel b: Distribution of distance error for Pareto optimal PCC circuits of size $\boldsymbol{n_{pos}=45, n_{neg}=39}$ (corresponding to the black line in Panel a). Area results are post-synthesis and normalized relative to the exact circuit.}%\vspace{-2ex}
    \label{fig:pcc_join}
\end{figure}

Using the methods outlined in Section~\ref{sec:ppc}, we generate $5,\!841$ approximate PCC circuits for our five target datasets.
Figure~\ref{fig:pcc_estimated} presents an area-versus-error analysis for three distinct PCC circuits customized for the Arrhythmia dataset.
The results reveal significant overlap in designs, indicating that the approximate PCC circuit design space can be substantially pruned.
This pruning reduces the size of the design space for TNN approximation.
Consequently, we limit the approximate PCC library to $134$ circuits across $22$ different input sizes.

Figure~\ref{fig:pcc_dist} presents a detailed error analysis of the design points on the black Pareto frontier from Figure ~\ref{fig:pcc_estimated}.
The logarithmic-scale histograms depict the distribution of distance error $D$ for $|G| = 10^6$ random input samples.
Hence, the x-axis represents the distance metric $D$ introduced in Section~\ref{sec:ppc}, while the y-axis shows normalized area results, with area savings increasing as you move down the axis.
The analyzed PCC circuit features $n_{pos}=45$, $n_{neg}=39$, and a $6$-bit comparator, corresponding to the top hidden-layer neuron of our bespoke Arrhythmia TNN model. 

Each row in the plot represents a Pareto-optimal approximation of the PCC circuit.
At the top, the fully accurate design's histogram is entirely concentrated in the bin for $D=0$, indicating no errors.
As we move down the rows, accuracy is traded for reduced area, causing the probability mass to gradually spread to bins with distances further from $0$.
If our exploration allows differences between positive and negative PC circuits up to $\varepsilon_{wcde}=4$, it results in: $12.6$\% area reduction (\nth{6} row from the top in Figure~\ref{fig:pcc_dist}), $95.57$\% error-free PCC operations, and a mean distance $\varepsilon_{mde}$ of only $0.06$.
Moving down the y-axis, for $\varepsilon_{wcde} \leq 8$, area savings increase to approximately $30$\%.
For approximations at the bottom of the plot, area can be reduced to as little as $25$\% of the accurate design while still maintaining over $80$\% correct results.
These results are consistent across all PCC circuit configurations examined.

\begin{figure}[t]
    \centering
    \includegraphics[width=\columnwidth]{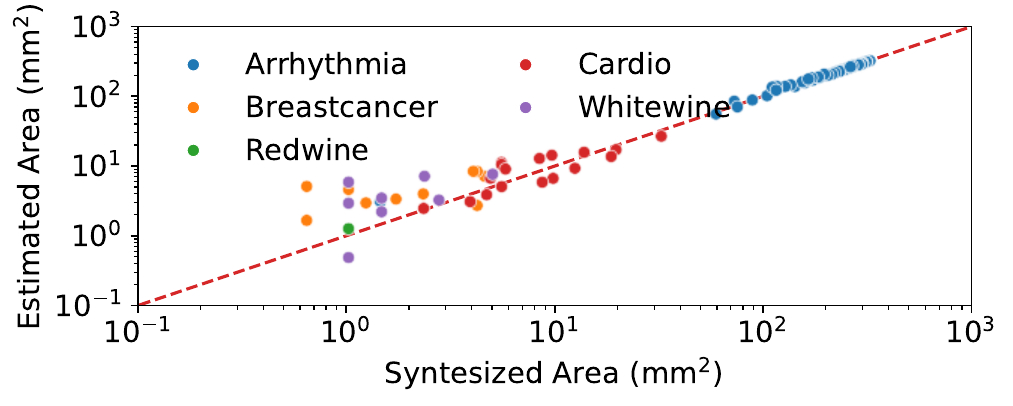}\vspace{-1em}
    \caption{Comparison between area estimates and post-synthesis results for various PCC circuits. Design points with similar colors indicate the use of the same exact design as a starting point.}
    \label{fig:pcc_area}%\vspace{-3ex}
\end{figure}

Given that the results presented in the first leg of Pareto analysis are based on area estimates, the final step is to evaluate how closely these estimates match the actual synthesis results.
\autoref{fig:pcc_area} provides this comparison, illustrating the relationship between estimated and synthesized area values for the PCC circuits.
While our estimation tends to underestimate smaller PCCs due to not accounting for comparator cost, occasional overestimations occur due to synthesis optimizations.
However, overall, there is a good enough correlation between the actual and estimated area.

\begin{figure*}[t]
    \centering
    \includegraphics[width=\textwidth]{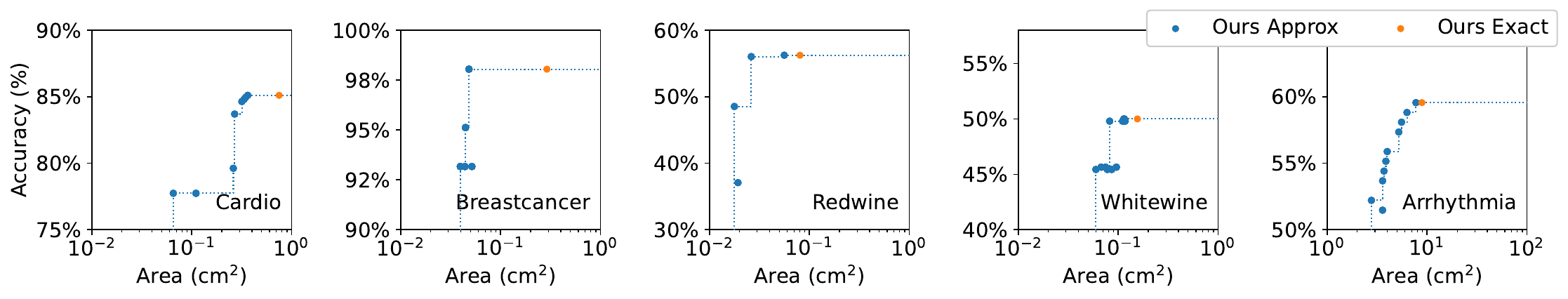}
    \caption{Post-synthesis area-versus-accuracy evaluation for Pareto-optimal approximate TNNs.%The orange points represent exact TNN circuit.
    }
    \label{fig:moo_pareto}
\end{figure*}

\begin{figure}[b]
    \centering
    \includegraphics[width=\columnwidth]{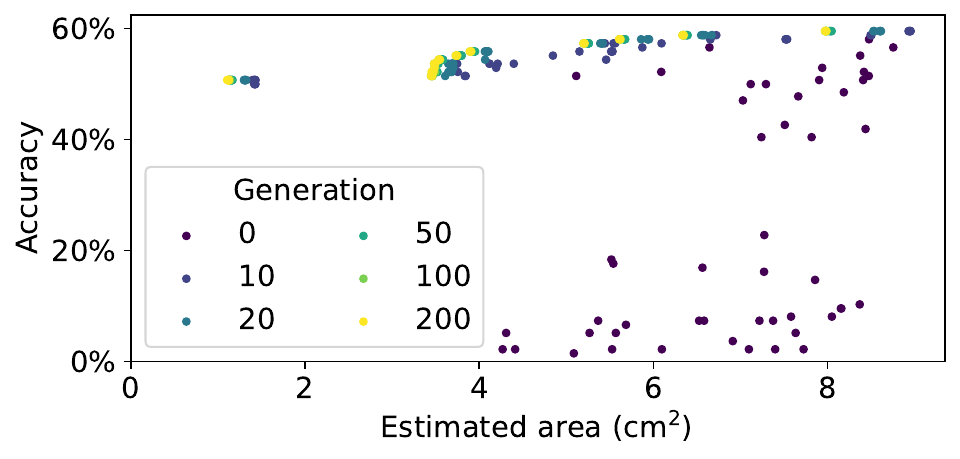}\vspace{-1em}
    \caption{Evolutionary optimization of the Arrhythmia TNN over 200 generations.}
    \label{fig:nsga:search}%\vspace{-2ex}
\end{figure}

\subsection{Approximate TNN Evaluation}
The final stage of our evolutionary approximation approach utilizes the NSGA-II algorithm to construct a TNN design.
This design incorporates components from our PCC (hidden neurons) and PC (output neurons) circuit libraries.
We implement NSGA-II using the \texttt{pymoo} library, employing an integer-based representation.
Each integer in this representation corresponds to the index of an approximate component for a given neuron.
Depending on the dataset under examination, chromosome lengths range from $6$ to $19$ integers. 

To navigate the design search space, which ranges from $2.4\cdot10^3$ approximate candidates for Cardio to $6.8\cdot10^{11}$ for Arrhythmia, we employ binary crossover with polynomial mutation.
If we were to consider PC approximation for every popcount operation in the hidden-layer neurons instead of combining them in Pareto-optimal PCC circuits, these values would increase to $5.2\cdot10^5$ and $1.7\cdot10^{14}$, respectively, complicating the required exploration.
The results of the design process for the Arrhythmia TNN (our largest TNN) are shown in~\autoref{fig:nsga:search}.
The algorithm was run for $200$ generations, taking less than $5$ minutes to complete, with substantial progress achieved within the first $50$ generations.

\autoref{fig:moo_pareto} presents the accuracy-versus-area results of our explorations, featuring designs found on the Pareto optimal curve.
Our findings reveal two key points: first, the generated approximate TNN designs can match the accuracy of their exact TNN counterparts while achieving an average area reduction of 41\%; second, if we allow for up to a 5\% decrease in accuracy compared to the exact TNN, the area savings increase to an average of 67\%.\smallskip

\noindent\textbf{Comparison with State-of-the-art:}

\begin{table}[t!]
\caption{Comparison against the State of the Art. All circuits target the EGFET technology.}\label{tab:soacomp}
\small\vspace{-1em}
\renewcommand{\arraystretch}{1.1}
\setlength\tabcolsep{2pt}
\begin{tabular}{ll|c|SS|SS}
\toprule
  &    &   & \multicolumn{2}{c|}{\textbf{w/o ADC cost}} & \multicolumn{2}{c}{\textbf{w/ ADC cost}} \\
\cline{4-7}
 \textbf{Dataset} & \textbf{Ref.} & {\thead{\textbf{Accuracy}\\ (\%)}} & {\thead{\textbf{Area} \\ (cm$^{2}$)}} & {\thead{\textbf{Power}\\ (mW)}} &  {\thead{\textbf{Area} \\ (cm$^{2}$)}}  &  {\thead{\textbf{Power}\\ (mW)}} \\
\hline
Arrhythmia & Exact MLP~\cite{Mubarik:MICRO:2020:printedml} & 62 & 266.00 & 998.00 & 298.06 & 1273.92 \\ 
Arrhythmia & Ax. MLP~\cite{Afentaki:ICCAD2023:axmac} & 60 & 13.51 & 12.80 & 45.57 & 288.72 \\ 
Arrhythmia & Our Exact TNN & 60 & 8.87 & 8.09 & 9.06 & 16.31 \\
Arrhythmia & Our Ax. TNN & 60 & 7.73 & 7.12 & 7.92 & 15.34 \\
Arrhythmia &  Our Ax. TNN & 57 & 5.2 & 4.95 & 5.39 & 13.17 \\

\hline 

BreastCancer & Exact MLP~\cite{Mubarik:MICRO:2020:printedml} & 98 & 12.00 & 40.00 & 13.17 & 50.07 \\ 
BreastCancer & Ax. MLP~\cite{Armeniakos:TC2023:codesign} & 97 & 5.10 & 18.00 & 6.27 & 28.07 \\ 
BreastCancer & Ax. MLP~\cite{Armeniakos:TC2023:codesign} & 93 & 1.50 & 5.69 & 2.67 & 15.76 \\ 
BreastCancer & Ax. MLP~\cite{Afentaki:ICCAD2023:axmac} & 94 & 0.08 & 0.08 & 1.25 & 10.15 \\ 
BreastCancer & Ax. MLP~\cite{Afentaki:DATE2024:gatrain} & 94 & 0.03 & 0.03 & 1.20 & 10.10 \\ 
BreastCancer & Our Exact TNN & 98 & 0.29 & 0.31 & 0.30 & 0.61 \\
BreastCancer & Our Ax. TNN & 98 & 0.05 & 0.04 & 0.06 & 0.34 \\
BreastCancer & Our Ax. TNN & 93 & 0.04 & 0.04 & 0.05 & 0.34 \\

\hline

Cardio & Exact MLP~\cite{Mubarik:MICRO:2020:printedml} & 88 & 33.40 & 124.20 & 35.86 & 145.35 \\ 
Cardio & Ax. MLP~\cite{Armeniakos:TC2023:codesign} & 87 & 6.08 & 20.80 & 8.54 & 41.95 \\ 
Cardio & Ax. MLP~\cite{Armeniakos:TC2023:codesign} & 83 & 4.92 & 16.80 & 7.38 & 37.95 \\ 
Cardio & Ax. MLP~\cite{Afentaki:ICCAD2023:axmac} & 85 & 1.35 & 1.57 & 3.81 & 22.72 \\ 
Cardio & Ax. MLP~\cite{Afentaki:DATE2024:gatrain} & 87 & 1.46 & 1.70 & 3.92 & 22.85 \\ 
Cardio & Our Exact TNN & 85 & 0.75 & 0.91 & 0.76 & 1.54  \\
Cardio & Our Ax. TNN & 85 & 0.36 & 0.42 &  0.37 & 1.05 \\
Cardio & Our Ax. TNN & 84 & 0.27 & 0.30 & 0.28 & 0.93  \\

\hline

RedWine & Exact MLP~\cite{Mubarik:MICRO:2020:printedml} & 56 & 17.60 & 73.50 & 18.89 & 84.58 \\ 
RedWine & Ax. MLP~\cite{Armeniakos:TC2023:codesign} & 55 & 1.06 & 3.95 & 2.35 & 15.02 \\ 
RedWine & Ax. MLP~\cite{Armeniakos:TC2023:codesign} & 51 & 0.87 & 3.25 & 2.16 & 14.33 \\ 
RedWine & Ax. MLP~\cite{Afentaki:ICCAD2023:axmac} & 55 & 0.03 & 0.02 & 1.32 & 11.10 \\ 
RedWine & Ax. MLP~\cite{Afentaki:DATE2024:gatrain} & 52 & 0.03 & 0.03 & 1.32 & 11.11 \\ 
RedWine & Our Exact TNN & 56 & 0.08 & 0.09 & 0.09 & 0.42   \\
%RedWine & Our Ax. TNN & 56 & 0.06 & 0.07 & 0.07 & 0.40  \\
RedWine & Our Ax. TNN & 56 & 0.03 & 0.03 & 0.04 & 0.36  \\

\hline

WhiteWine & Exact MLP~\cite{Mubarik:MICRO:2020:printedml} & 54 & 31.20 & 126.40 & 32.49 & 137.48 \\ 
WhiteWine & Ax. MLP~\cite{Armeniakos:TC2023:codesign} & 53 & 6.47 & 21.30 & 7.76 & 32.38 \\ 
WhiteWine & Ax. MLP~\cite{Armeniakos:TC2023:codesign} & 49 & 0.79 & 2.77 & 2.07 & 13.84 \\ 
WhiteWine & Ax. MLP~\cite{Afentaki:ICCAD2023:axmac} & 50 & 0.25 & 0.25 & 1.54 & 11.33 \\ 
WhiteWine & Ax. MLP~\cite{Afentaki:DATE2024:gatrain} & 51 & 0.23 & 0.25 & 1.52 & 11.33 \\ 
WhiteWine & Our Exact TNN & 50 & 0.16 & 0.18 & 0.17 & 0.51  \\
WhiteWine & Our Ax. TNN & 50 & 0.11 & 0.12 & 0.12 & 0.45 \\
WhiteWine & Our Ax. TNN & 45 & 0.06 & 0.07 & 0.07 & 0.40  \\

\bottomrule
\end{tabular}

\end{table}

\autoref{tab:soacomp} presents a comparison of our results against the current state-of-the-art in printed neural networks.
Our approximate TNN designs demonstrate a superior accuracy-area trade-off compared to approximate multilayer perceptrons, even without considering the sensor-processor interface cost.
Within a $1$\% accuracy variation, our TNNs achieve, on average, $42\times$, $2.17\times$, and $1.99\times$ lower area compared to previous works on approximate multilayer perceptrons~\cite{Armeniakos:TC2023:codesign, Afentaki:ICCAD2023:axmac, Afentaki:DATE2024:gatrain}.
Power consumption is similarly reduced by $162\times$, $2.05\times$, and $1.97\times$, on average, respectively.
When sensor-processor interface costs (ADCs and ABCs) are factored in, these gains grow further, with area reductions exceeding $6\times$ and power savings of over $19\times$ compared to the most efficient reported multilayer perceptron design~\cite{Afentaki:DATE2024:gatrain}.
When compared to an exact multilayer perceptron implementation~\cite{Mubarik:MICRO:2020:printedml}, our approximate TNNs achieve $55\times$ lower area and $97\times$ less power at the expense of a $5$\% accuracy drop. 

Notice that all our TNN circuits, except those targeting the Arrhythmia dataset, consume well below $2$~mW, even when considering the power consumed by the ABCs and the power overhead of the sensors (approximately $5$~$\mu$W based on previous studies~\cite{Bleier:ISCA:2020:printedmicro}).
Thus, they can be powered by a printed energy harvester~\cite{printedharvester}.
To the best of our knowledge, this is the first digital printed classifier to achieve this.
As for our Arrhythmia TNN designs, they can be powered by either a Zinergy ($15$~mW) or Molex ($30$~mW) printed battery~\cite{Mubarik:MICRO:2020:printedml}.
This represents a significant advancement over previous solutions~\cite{Mubarik:MICRO:2020:printedml, Afentaki:ICCAD2023:axmac} targeting this application that exceeding the power budget of printed batteries.

\section{Conclusion}
Printed electronics, particularly classifier circuits, offer transformative potential in applications requiring flexible substrates and ultra-low costs.
Despite recent advancements, practical implementation faces challenges due to low integration density and limited power from printed batteries and harvesters.
Our research approaches these issues holistically, focusing on optimizations from the sensor-processor interface to the processor itself.
To our knowledge, we present the first open-source end-to-end digital printed classifier that meets both resource and power constraints
Key contributions include: replacing ADCs with more efficient ABCs, customizable to specific input features; implementing bespoke ternary neurons, without costly multipliers and with approximate popcount and popcount-compare units; developing and deploying a multi-stage evolutionary optimization method that assists both in the approximation of the popcount and popcount-compare circuits and in their integration into a complete TNN accelerators.
When sensor-processor interface costs are factored in, our results show a 32$\times$ improvement in area and 34$\times$ reduction in power compared to current state-of-the-art approximate digital printed neural network implementations, while maintaining comparable accuracy across diverse datasets.

% In this study, we focus on the design of printed Neural Networks, with a specific consideration for the hardware cost of ADCs, aiming at facilitating sensor-based printed applications.
% To address this challenge, we introduce a printed customizable analog-to-binary converter with learnable thresholds per input, enabling the utilization of binary inputs with our TNNs.
% To overcome the stringent hardware constraints inherent in printed electronics, we implement a holistic approximation approach tailored to our TNN circuits.
% We establish novel libraries of approximate popcount and popcount-compare units, leveraging CGP and BDD formal verification methods, and exploiting a novel error metric for the design of the latter. 
% Through our evolutionary multi-objective optimization framework, we effectively navigate the complex trade-offs between accuracy and hardware cost in TNN design, enabling their operation using printed energy-harvesters in all but one of the examined cases.

% \todo{ Georgios?
% Outcomes
% - new library of PC build using CGP and BDD formal verification
% - we proposed a new distance metric for approximate comparing operators
% - in the design we handle unconventional EGFET technology
% - we involved multiobjective optimization to obtain tradeoffs between accuracy and hw cost of TNNs
% }

\begin{acks}
This work was supported by the Czech Science Foundation project
22-02067S, by the European Research Council (ERC), and by the H.F.R.I call “Basic research Financing (Horizontal support of all Sciences)” under the National Recovery and Resilience Plan “Greece 2.0” (H.F.R.I. Project Number: 17048).
The authors thank Ampere for server support.
\end{acks}
%\newpage


\bibliographystyle{ACM-Reference-Format}
\bibliography{iccad}

\end{document}